\documentstyle[preprint,aps]{revtex}
\begin{document}

\draft

\title{Charge Density Wave
Behaviour of the Integer Quantum Hall Effect Edge States.}

\author{M. Franco and L. Brey}
\address{Instituto de Ciencia de Materiales (CSIC).
Universidad Aut\'onoma de Madrid C-12,
Cantoblanco, 28049, Madrid, Spain.  }
\date{\today}

\maketitle

\begin{abstract}
We analyze the effect that the Coulomb interaction has on the edge excitations
of an electron gas confined in a bar of thickness $W$,
and in presence of a magnetic field corresponding to
filling factor 1 Quantum Hall effect.
We  find that the long-range interaction between the edges leads
the system to a  ground state with a quasi-long range order, similar
to a Charge Density Wave.
The spectral density of states vanishes  at zero frequency, and increases
with frequency faster than any power law, being the conductance of
a infinite long system zero.

\end{abstract}
PACS number 73.40.Hm, 72.15.Nj

Edge states play a very important role in the understanding of the
Quantum Hall Effect (QHE).\cite{books}
Some properties of the integer QHE can be understood from the one-electron
picture of the edge states\cite{butt}, but at the moment
the edge structure in the fractional QHE regime is unclear.\cite{edfr}
Due to the one-dimensional (1D) character of the QHE edge states,
it has been proposed that they behave as a Fermi liquid in the integer
case and as a Luttinger liquid in the fractional case.\cite{wen1,stone}
In these works, it is assumed short range electron-electron interaction and
therefore it is also neglected the interaction between outgoing and ingoing
edge states.
However, in order to obtain high
mobility electron samples, any metallic gate in the actual physical
systems must be far
away from
the plane where the two-dimensional electron gas (2DEG) lives, and therefore
the
Coulomb
interaction between electrons is practically unscreened .

In this paper we  study the effect that the Coulomb interaction
has on the properties of the edge states of an electron gas confined in a
bar of thickness $W$, in the case  of QHE at   filling factor 1.
Our main conclusion  is that the long-range Coulomb interaction between
edges leads the system to a state characterized by a quasi-long-range order,
similar to an 1D charge density wave (CDW) state.  We obtain
that
the charge-charge correlation function along the edge direction ($y$)  presents
oscillations with period
$2 \pi /W$ which decay slower than any power law of $y$.
(Along this work we take the magnetic length, $\ell$, as the unit length and
 $ e ^ 2/\ell$ as the unit of energy).
We find that the the spectral density of states is zero at zero frequency
($\omega$),
and increases with  $\omega$ faster than any power law.
Also we find that this system has not a well defined Fermi surface, although
the derivate of the ocupations at wavevector $W/2$ tends to infinity with
the thickness of the electron bar.
Due to the quasi-long-range order, the conductance of a infinity long system
is zero, but
we find that for realistic finite
systems it depends strongly on the length and thickness of the bar.

 Since we are dealing with a strong magnetic field ($B$) phenomenon, and we are
interested in edge excitations with  smaller energy than the Zeeman and
cyclotron energy, we only consider states in the lowest Landau level
and with the spin parallel to B. In the Landau gauge the Hamiltonian has the
form,
\begin{equation}
H= \sum _ {k} \epsilon _0 (k) c _k  ^{+} c _ k
+ {1 \over {2 L _y L _x}} \sum _{k,k',\vec{q}} v(q)
e ^ {- q ^ 2 / 2} e ^ { i q _x ( k - k '+q _ y)}
c _ k ^ {+} c _ {k'} ^ {+}
c _ {k' + q _ y} c _{k - q_y} \, \, .
\end{equation}
Here $c_k ^ + $ creates an electron with wave function
${1 \over {\sqrt{ L_y \pi ^{1/2}}}} e ^{i k y} e ^{ - (x-k)^2/2}$, $L_x$ and
$L_y$
are the system dimensions and $v(q)$ is the Fourier transform of the Coulomb
interaction. The first term in equation (1) is the confinement potential and
it is choosen to be the created by a fictitious positive background which
cancels the electronic charge corresponding to get occupied all the electron
states
with $-W/2\!<\!k\!<\!W/2$.
Assuming translational invariance in the $y$ direction,
the Hartree-Fock (HF) solution of this Hamiltonian is
\begin{equation}
|\Psi _ {HF}>= \prod _{-W/2 < k < W/2} c _k ^ + |\emptyset>\, \, \, .
\end{equation}
This solution coincides with the exact solution for large values
of $W$.\cite{chamon,ctp1}
The charge density excitations (CDE) of the ground state described by
$ |\Psi _
{HF}> $, have the form \cite{ctp1}
\begin{equation}
|\Psi _ q ^ i > = \sum _ k \alpha _ q ^ i (k ) c _ {k+q} ^+ c _ k
|\Psi _ {HF}> \, \, \, ,
\end{equation}
where the coefficients $ \alpha _ q ^ i (k ) $
are obtained by minimizing the energy of the CDE,
$\hbar \omega _ i (q)\!= <\Psi _ q ^ i|H|\Psi _q ^ i>
\! / \! <\Psi _ q ^ i|\Psi _ q ^ i>$i,
with respect them.
Due to the 1D character of these excitations, for a given $q$,
the number of CDE is given by $q L_y / 2 \pi$.
In figure 1 it is plotted the energy of the CDE versus
the wavevector $q$ for two different values of $W$.
The CDE spectrum consists of a continuous band of energy,
which increases in size with $q$, and some lower energy branches. The
continuum of energies corresponds basically to electron-hole excitations
and the low energy branches to collective excitations.
Using the HF wavefunction as the ground state,
our method of calculation of the CDE  energies is
equivalent to the time-dependent HF method, and we can
classified the different terms contributing
to the energy of the CDE as:
(a) a term corresponding to the confinement potential,
(b) a term coming from the self-energies
difference between the excited electron and the level from which the
electron is removed,
(c) a depolarization term coming from terms in the
Hamiltonian where the electron and hole annihilate each other at one point
in space and one electron-hole pair is created simultaneoulsly at onother
point, and
(d) and excitonic term  which is just the direct Coulomb
interaction of the excited electron and hole.
At small $q$ the dispersion of the collective excitations
has the form $
\hbar \omega _ i (q) = ( A _ i - {1 \over \pi} \ln {q}) q
$, where
the coefficients $A _i$ depend only on the confinement potential and the
$q \ln {q}$ term corresponds to the classical dispersion of edge
magnetoplasmons\cite{volkov,wassermeier}
and comes from the depolarization term.
It is important to note
that at small $q$ there is an exact cancelation between the contributions
(b) and (c) to the energies of the CDE, in such a way that
the expected classical result, coming from the depolarization term, is
obtained.

The lowest energy CDE always shows a deep minimum at $q \sim W$. This
minimum is due to the coupling between the two edges of the system and it
becomes a soft mode when $W$ is smaller than 4 magnetic lengths.
The existence of this soft mode is an indication  of the importance that
quantum fluctuations have on this system when $W$ is small. It is necessary
to go beyond mean field approximation in order to describe the properties of
the edge states. The idea is to write an effective Hamiltonian, $H_
{eff}$, for describing the low-energy excitations of the system, i.e.
the edge excitations. It is not possible to write an $H_{eff}$ as a function
of the  CDE previously obtained because they are not  independent between
them. However, due to the 1D character of the edge excitations they can be
generated by repeated application of the density operators on the ground
state.\cite{Haldane} This is strictly true for the Luttinger
model\cite{Haldane,lutt,Mahan},
and for physical systems  it is based on the assumption that the following
conmutation relation are satisfied:
\begin{eqnarray}
& [ & \rho _ 1 (q), \rho _ 1 (-q')]= q {{L_y}\over {2 \pi}} \delta _{q,q'}
\nonumber
\\
& [ & \rho _ 2 (q) , \rho _ 2 (-q')]= - q {{L_y}\over {2 \pi}} \delta _{q,q'}
\nonumber
\\
& [ & \rho _1 (q), \rho _2 (q')]= 0 \, \, \, ,
\end{eqnarray}
with $\rho _1 (q)\!\! =\! \!\sum _{k>0} \!\! :\! \!c _ {k+q} ^ + c _ k\!\! :$
and $\rho _2 (q)\!\! =\!\!
\sum _{k<0}\! \! :\! \!c _ {k+q} ^ + c _ {k}\! :\!\! $. Here, the colons
represent normal ordering of the creation and annihilation operators.
For $q>0$, the operator $\rho _1 (q)$ ($\rho _2 (-q)$) acting of the
groundstate,
creates an excitation on
the right (left) edge of the bar.
Although expressions (4)  are not strictly satisfied in our system, its
average in the HF ground state is satisfied for $q< W$, and errors in using
equations (4) are
proportional to $q$, therefore we expect that using commutation
relation(4) is a good approximation for small $q$.

Using  the operators $\rho _ 1$ and $ \rho _ 2$ as the generators of the low
energy excitations in the system (edge excitations), the effective
Hamiltonian can be written, in the harmonic approximation, as
\begin{equation}
H_{eff}=  \sum _ {q >0}  { {2 \pi } \over { q L _y}} \left (
\hbar \omega ( q)
\left  [ \rho _1 (-q) \rho _1 (q) + \rho _ 2(q) \rho _ 2 (-q) \right ]
+ V (q) \left  [
\rho _ 1 ( -q) \rho _ 2 (q) + \rho _ 2 (-q) \rho _1 (q) \right ] \right )
\, \, \, ,
\end{equation}
with
\begin{eqnarray}
& \hbar  & \omega ( q ) = <\Psi _ {HF} | \rho _ 1 (-q) H \rho _ 1 (q) |\Psi
_{HF}> \, \, \, \,  {\rm and}  \nonumber \\
& V &(q)=<\Psi _{HF} |H \rho _1 (-q) \rho _2 (q)|\Psi _ {HF}>\, \, \, .
\end{eqnarray}
Two terms contribute to $V(q)$, a depolarization and a excitonic term.
We find, that due to the short range of the exchange interaction, for $W$
larger than 2 magnetic lengths the excitonic term is practically zero and
the depolarization term get its classical value
$V(q)={q \over \pi} K_0(q W)$.
The value of $\omega (q)$ is practically equal to the lower energy CDE
$\omega _ 1 (q)$ for wavevectors $q<1$ and $W>2$.

The Hamiltonian (5) can be solved using a Bogoliubov
transformation\cite{mattis} and the elementary excitations have energy
$E_q=\sqrt {
\hbar ^ 2 \omega _ q ^ 2 - V _ q ^ 2 }$.
Due to the coupling between the edges, at small wavevectors
($q << W^{-1}$)  $E_q$ disperses as a
1D plasmon:
$ E _q \sim q  \sqrt {
|\ln {q}|}$.
These excitations can still be classified as left and right movers,
but now a mode propagating on an edge is dragged by the charge in the
other edge.\cite{orgad}
{}From the solution of the effective Hamiltonian (5)
it is possible to obtain\cite{Mahan}
the left and right one electron creation operators,
and from them to compute different correlation functions.
The two Fermi wavevector ($W$) CDW correlation function can be
written as follows
\begin{equation}
<\!\rho (y) \rho (0)\!> _{W} =\!  \lim _ {\alpha \rightarrow 0}
{ 1 \over { \pi \alpha}}  \cos {(Wy) } \exp  \left [
{  2 \int _0 ^{\infty}
{{ e ^{ -\alpha q} } \over q }
{ { \hbar \omega (q) - V (q) } \over { E _ q}}(\cos {q y } -1)}
\right ] \, \, \, .
\end{equation}
We find that for large values of  $y$,
$<\!\rho (y) \rho (0)\!> _{W}$  decays as
$ e ^ { -const  (\ln{y} \ln{W}) ^ {1/2} } $, i.e. slower than
any power law.
The slow decay with $y$ of
this CDW correlation function indicates an
incipient CDW
ground state with wavevector $W$, in which the chage density waves in the
left and right
edges are out of phase.
This result is very similar to the one obtained by
Schulz\cite{schulz} for the case of an 1D electron gas with long range
Coulomb interaction.
However, in our case, due to the spatial separation between the outgoing
and ingoing edges, the period of the oscillations in the
density  density correlation function, is related with the inverse of the
lineal density of charge in the bar ($\sim W$), and not with the average
edge interparticle spacing ($\sim \ell$).

Using the bosonization method we also calculate the single particle Green
function, which for large values of $ y$ decays as
$ e ^ { -const ( \ln {y }) ^ {3/2} / (\ln {W}) ^{1/2}} $, i.e.
much faster than any power law.\cite{schulz}
{}From the single particle Green function we compute the momentum distribution
function $< \!n _ k \! >$ for different values of $W$. In figure 2 we plot
$ <\!n_k \!>$
versus $k$ for two values of $W$. Due to the fast decay of the one particle
Green function, $< \! n_k \!>$ and all
its derivates are continuos at the Fermi
wavevector of the system $W/2$. Nevertheless, we can see in figure 2
that the absolute  value of the
derivate of $< \! n_k \! >$ at $W/2$ increases  very quicly with  $ W $.
Therefore for large values of $W$, although
the system does have not a Fermi surface, $<\! n_k \!>$ changes from 1 to 0
very
abruptly at $W/2$.

{}From the retarded current current correlation function
we have also studied
the conductivity of the system, which in this case has the form,
\begin{equation}
\sigma ( q, \omega) =  \lim _{\delta \rightarrow 0} {{2 e ^ 2} \over h}
{ { \omega _ q - V_q /\hbar} \over q}
{ { i ( \omega + i \delta) } \over
{ ( \omega + i \delta ) ^ 2 - ( E _q/\hbar) ^ 2 } } \, \, .
\end{equation}
{}From the conductivity it is possible to calculate the conductance,
\begin{equation}
G= \lim _{ q \rightarrow 0} { { e ^ 2} \over { h}} \, { {2 \pi} \over {L _y}}
 { { \omega _ q - V_q / \hbar} \over {q}}  \delta ( E_q/\hbar) \, \, ,
\end{equation}
which is zero for  the case of an infinite system. This result is a consequence
of the quasi long range order existing in the system.
Only in the $W\rightarrow \infty$ case the conductance in the system
corresponds to the case of an isolated edge, $ e^ 2 /h$.
For a better  understanding of this result we calculate the
momentum-integrated spectral density function,
\begin{equation}
A(\omega)=
\lim _{ \alpha \rightarrow \infty} { 1 \over { \pi \alpha}}
\int _{-\infty} ^{\infty} dt
e ^{ i \omega t}
\exp { \left [
- \int _ 0 ^{\infty} dp  { {e^{- \alpha p}} \over p}
{{\hbar \omega (p)}  \over {E_p}}
( 1 - e ^ {-i E _p t} )  \right ]} \, \, ,
\end{equation}
which  satisfies the numerically more convenient  integral equation
\begin{equation}
A(\omega) \omega = \int _0 ^{p _c} dp {{ \omega (p) } \over p }
A(\omega - E _p) \, \, \, ,
\end{equation}
with $p_c$ verifying  $E_{p_c}= \hbar \omega$.
For small frequencies  $A(\omega) \sim e^ {-const ( - \ln( \omega)) ^{3/2} /
(\ln W )^{1/2} }$.
In a Luttinger liquid with conductance $G ^{lutt} = g { {e ^2} \over h}$,
the spectral density of states varies as
$ A ^{lutt} (\omega)= \omega ^{(g+1/g)/2 -1}$, therefore as smaller is the
conductivity bigger is the power in the spectral density.\cite{apel,kane}
In our case the density of states increases with $\omega$ faster than any
power law of $\omega$, and it is in agreement with the fact that
the conductance is zero.
On the contrary, in the case of not coupling between edges ($V(q)=0$),
the calculation reduces to the HF approximation for isolated edges, and
the density of states vanishes at the Fermi energy, increasing with the
frequecy as
$(-\ln ( \omega))^{-1}$, i.e. slower than
any power law, being therefore the conductance $e ^ 2 / h$.

Since the conductance goes to zero as $1/ (- \ln q) ^{1/2}$ it
is convenient a
estimation of the conductance in a  large but finite system.
In figure 3 we plot, for different values of $W$, the quantity
\begin{equation}
K(q,W)= { { \omega _q - V _ q / \hbar } \over {q}} \,
\left ( { { \partial E _ q } \over { \hbar \partial q }  } \right ) ^{-1}  \,
\, ,
\end{equation}
as a function of $(q W ) ^ {-1}$. For large values of $L_y/W$, the
conductance is $G\sim {  { e ^2} \over h }K( {{ 2 \pi} \over { L _ y }}, W )$,
and from figure 3 we see that for realistic  large
values of $L_y$, the
conductance is different from zero. Also we  see in figure 3 that,
due to the long range of the Coulomb interaction,
$G <  e^2 /h$ even for very large values of $W$.

We should stress here that the conductance we compute corresponds to the
two terminal conductance and not to the Hall conductance. In fact, if the
density of states at the Fermi energy is not singular, the
Hall conductance has the value $e^2 / h$\cite{oreg}, independently
of the coupling between the edges.
Thus the Hall transport does not provide information on the edge interactions.
We propose that the two-terminal conductance could give us  information
on the interaction between electrons in different edges.
Of course, experimentally the measure of the two-terminal conductance
can be influenced by the ohmic constact resistance,
and it could be difficult to separate the correction to the
conductance due to  edge coupling from  the one
due to contacts.

In summary, we have studied the effect that the Colulomb interaction
has on the properties of the integer QHE edge states,
of a Hall bar as a function  of  its thickness $W$.
We find that the system is characterized by a quasi-long range order,
similar to a CDW. The spectral density of states vanishes at
the Fermi energy and increases with $\omega$ slower than any power law,
and therefore the two terminal conductance of a infinite system should
be zero.
However we estimate that for large but finite system the   values of
the conductance  depend on the thickness and length of the Hall bar, and
is not quantized.

We are thankful to G.G\'omez Santos, F.Guinea, A.H.MacDonald, C.Tejedor and
I.Zapata for discussions on the subject and to L.Martin-Moreno for a
critical reading oft he manuscript.  This work has been supported in
part by CICyT of Spain under contract No. MAT94-0982. L.B. was supported by
a NATO fellowship during part of the realization of this
work.

\begin{figure}
\caption {
Dispersion relation,
obtained in the time dependent Hartree-Fock approximation,
of the lowest energy charge density excitations
for  (a) $W=10 \ell$ and (b) $W =2\ell$.
}
\label{fig1}
\end{figure}
\begin{figure}
\caption {
Variation as a function of $\omega$ of the absolute value of the derivate of
the
momentum distribution function,
$<n_k>$ evaluated at the Fermi wavevector $W/2$.
In the inset of the figure it is plotted $<n _k>$ as a function
of $k$ for two different values of $W$.
}
\label{fig2}
\end{figure}
\begin{figure}
\caption {
Variation of the quantity $K(q,\omega)$, see text, as a function of $qW$,
for different values of $W$. }
\label{fig3}
\end{figure}
\end{document}